\documentstyle[amsfonts,epsfig,12pt]{article}

\input{tcilatex}

\begin{document}

\bigskip \bigskip 
\begin{titlepage}
\bigskip \begin{flushright}
WATPHYS-TH01/08\\
hep-th/0111217
\end{flushright}


\vspace{1cm}

\begin{center}
{\Large \bf {Action, Mass and Entropy of  Schwarzschild-de Sitter black holes  and the 
de Sitter/CFT Correspondence    }}\\
\end{center}
\vspace{2cm}
\begin{center}
A. M. Ghezelbash{%
\footnote{%
EMail: amasoud@sciborg.uwaterloo.ca;  On leave from Department of Physics,
Az-zahra University, Tehran, Iran}} and
R. B. Mann
\footnote{
EMail: mann@avatar.uwaterloo.ca} \\
Department of Physics, University of Waterloo, \\
Waterloo, Ontario N2L 3G1, CANADA\\
\vspace{1cm}
\today\\
\end{center}

\begin{abstract}
We investigate a recent proposal for defining a conserved mass in
asymptotically de Sitter spacetimes that is based on a conjectured 
holographic duality between such spacetimes and Euclidean conformal
field theory.    We show that an algorithm for deriving such terms
in asymptotically anti de Sitter spacetimes has an asymptotically de Sitter
counterpart, and derive the explicit form for such terms up to 9 dimensions.
We show that divergences 
of the on-shell action for de Sitter spacetime are removed in
any dimension in inflationary coordinates, but in covering coordinates 
a linear divergence remains in odd dimensions that cannot be cancelled by 
local terms that are polynomial in boundary curvature invariants.
We show that the class of  Schwarzschild-de Sitter black holes up to
9 dimensions has finite action and conserved mass, and construct a 
definition of entropy outside the cosmological horizon 
by generalizing the Gibbs-Duhem relation in 
asymptotically dS spacetimes.  The entropy is agreement with that
obtained from CFT methods in $d=2$. In general our results provide further
supporting evidence for  a dS/CFT correspondence, although some 
important interpretive problems remain.
\end{abstract}
\end{titlepage}\onecolumn      

\section{Introduction}

It is generally believed that the definition of a conserved charge in a
gravitational spacetime that is asymptotically de Sitter (dS) is not well
defined. The reason is that such spacetimes do not have spatial infinity the
way that their asymptotically flat or asymptotically anti de Sitter (AdS)
counterparts do. Moreover one cannot define a timelike Killing vector in
global de Sitter spacetime. In fact, there is a timelike Killing vector
field inside the cosmological horizon that becomes spacelike outside the
cosmological horizon. For this reason, the physical meaning of the \
Abbott-Deser energy outside the cosmological horizon of dS spacetime is not
clear and to construct the energy, one could use the conformal Killing
vector \cite{Shiro}.

Recently a novel prescription was proposed for computing conserved charges
(and associated boundary stress tensors) of asymptotically dS spacetimes
from data at early or late time infinity \cite{bala}. The method is
analogous to the Brown-York prescription in asymptotically AdS spacetimes %
\cite{brown,BCM,balakraus}, and yields suggestive information about the
stress tensor and conserved charges of the hypothetical dual Euclidean
conformal field theory (CFT) on the spacelike boundary of the asymptotically
dS spacetime, providing intriguing evidence for a holographic dual to dS
spacetime that is similar to the AdS/CFT correspondence. Such a similarity
also is observed in the computation of the conformal anomaly of dual
Euclidean conformal field theory \cite{serg}. The specific prescription in
ref. \cite{bala} (which has been employed previously by others but in more
restricted contexts \cite{Klem, Myung}) presented the counterterms on
spatial boundaries at early and late times that yield a finite action for
asymptotically dS spacetimes in $3,4,5$ dimensions. By carrying out a
procedure analogous to that in the AdS case \cite{BCM,balakraus}, one could
get the boundary stress tensor on the spacetime boundary, and consequently a
conserved charge interpreted as the mass of the asymptotically dS spacetime
could be calculated. Sample calculations led the authors of \cite{bala} to
the following conjecture: {\it Any asymptotically dS spacetime with mass
greater than dS has a cosmological singularity. \ }Although an exact proof
of this conjecture has not been attained, it has been verified for
topological dS solutions and its dilatonic variants \cite{cai}.

The purpose of this paper is to investigate this prescription in greater
detail. We first demonstrate in sections 2 and 3\thinspace\ that the
procedure for deriving boundary counterterms from the Gauss-Codacci equation
for asymptotically AdS spacetimes \cite{KLS} applies also to the
asymptotically dS case. We show that these counterterms are sufficient for
obtaining a finite action for the inflationary patches (big bang and big
crunch patches) of dS\ spacetime in any dimensionality. However such actions
are {\it not }finite when computing for the full dS spacetime using covering
coordinates: they contain a linear divergence in spacetimes of odd
dimensionality. This divergence is similar to that found in the AdS case %
\cite{EJM}.\ We then move on in section 4 to compute the action for a
Schwarzschild-de Sitter (SdS) black hole with dimensionality up to nine. \
We also compute the boundary stress tensor and mass of these SdS black
holes. \ We then define a notion of entropy outside of the horizon by
generalizing the gravitational Gibbs-Duhem relation to this situation. \ By
appropriately identifying a spatial coordinate outside of the horizon,
infinite volume divergences due to integration over this coordinate on the
boundary are removed, and our definition of entropy agrees with that
obtained using CFT methods in $3$ dimensions \cite{bala}. \ However the
justification and interpretation of these results and the above conjecture
is less than clear. For example masses greater than that of pure de Sitter
spacetime can be obtained by reversing the sign of the mass parameter,
whilst keeping all singularities hidden from observers outside of the
cosmological horizon. We comment on this in the final section.

\bigskip

\section{Boundary Counterterms}

In $d+1$ dimensions, the Einstein equations of motion with a positive
cosmological constant can be derived from the action 
\begin{equation}
S=I_{B}+I_{\partial B}  \label{action}
\end{equation}

where 
\begin{eqnarray}
I_{B} &=&\frac{\alpha }{16\pi G}\int_{{\cal M}}d^{d+1}x\sqrt{-g}\left(
R-2\Lambda +{\cal L}_{M}\right)  \label{actbulk} \\
I_{\partial B} &=&\frac{\beta }{16\pi G}\int_{{\cal \partial M}^{-}}^{{\cal %
\partial M}^{+}}d^{d}x\sqrt{h^{\pm }}K^{\pm }  \label{actbound}
\end{eqnarray}
and ${\cal L}_{M}$ refers to the matter Lagrangian, which we shall not
consider here.

The first term in (\ref{action}) is the bulk action over the $d+1$
dimensional Manifold ${\cal M}$ with Newtonian constant $G$ and the second
term (\ref{actbound}) is the Gibbons-Hawking surface term which is a
necessary term to ensure a well defined Euler-Lagrange variation. ${\cal %
\partial M}^{\pm }$ are spatial Euclidean boundaries at early and late times
and $\int_{{\cal \partial M}^{-}}^{{\cal \partial M}^{+}}d^{d}x$ indicates
an integral over the late time boundary minus an integral over the early
time boundary. The quantities $g_{\mu \nu },h_{\mu \nu }^{\pm }$ and $K^{\pm
}$ are the bulk spacetime metric, induced boundary metrics and the trace of
extrinsic curvatures of the boundaries respectively. We shall usually
suppress the ``$\pm $'' notation when it is obvious. For a well-defined
variational principle, we must have $\beta =-2\alpha $, and one typically
chooses $\alpha =1$ as an overall normalization\footnote{%
Our conventions are the same as in ref. \cite{KLS}.}. However, as is well
known the action (\ref{action}) is not finite when evaluated on a solution
of the equations of motion. The reason is the infinite volume of the
spacetime at early and late times.

\bigskip

The procedure for dealing with such divergences in asymptotically flat/AdS
cases (where they are large-distance effects) was to include a reference
action term \cite{brown,BCM}, which corresponded to the action of embedding
the boundary hypersurface ${\cal \partial M}$ (whose unit normal is
spacelike) into some other manifold. The physical interpretation is that one
has a collection of observers located on the closed manifold ${\cal \partial
M}$, and that the physical quantities they measure (energy, angular
momentum, etc.) are those contained within this closed manifold relative to
those of some reference spacetime (regarded as the ground state) in which $%
{\cal \partial M}$ is embedded \cite{nonortho}. For example in an
asymptotically anti de Sitter spacetime, it would be natural to take pure
AdS as the ground state reference manifold. \ 

However this procedure suffers from several drawbacks: the reference
spacetime in general cannot be uniquely chosen \cite{CCM} nor can an
arbitrary boundary ${\cal \partial M}$ always be embedded in a given
reference spacetime. Employing approximate embeddings can lead to ambiguity,
confusion and incompleteness; examples of this include the Kerr \cite%
{martinez}, Taub-NUT and Taub-bolt spacetimes \cite{TNUT}.

An alternative approach for asymptotically AdS spacetimes was suggested a
few years ago that has enjoyed a greater measure of success \cite%
{balakraus,hensken,korea,sergey}. It involves adding to the action terms
that depend only on curvature invariants that are functionals of the
intrinsic boundary geometry. \ Such terms cannot alter the equations of
motion and, since they are divergent, offer the possibility of removing
divergences that arise in the action (\ref{action}) provided the
coefficients of the allowed curvature invariants are correctly chosen. \ No
embedding spacetime is required, and computations of the action and
conserved charges yield unambiguous finite values that are intrinsic to the
spacetime. \ This has been explicitly verified for the full range of type-D
asymptotically AdS spacetimes, including Schwarzchild-AdS, Kerr-AdS,
Taub-NUT-AdS, Taub-bolt-AdS, and\ Taub-bolt-Kerr-AdS \cite%
{EJM,misner,nutkerr}. \ 

The boundary counterterm action is universal, and a straightforward
algorithm has been constructed for generating it \cite{KLS}. \ The procedure
involves rewriting the Einstein equations in Gauss-Codacci form, and then
solving them in terms of the extrinsic curvature functional $\Pi
_{ab}=K_{ab}-Kh_{ab}$ of the boundary ${\cal \partial M}$ and its normal
derivatives to obtain the divergent parts. \ It succeeds because all
divergent parts can be expressed in terms of intrinsic boundary data, and do
not depend on normal derivatives \cite{Feffgraham}. \ \ By writing the
divergent part $\tilde{\Pi}_{ab}$ as a power series in the inverse
cosmological constant \ the entire divergent structure can be covariantly
isolated for any given boundary dimension $d$; by varying the boundary
metric under a Weyl transformation, it is straightforward to show that the
trace $\tilde{\Pi}$ is proportional to the divergent boundary counterterm
Lagrangian.

Explicit calculations have demonstrated that finite values for the action
and conserved charges can be unambiguously computed up to $d=8$ for the
class of Kerr-AdS metrics \cite{saurya}. \ The removal of divergences is
completely analogous to that which takes place in quantum field theory by
adding counterterms which are finite polynomials in the fields. The AdS/CFT
correspondence conjecture asserts that these procedures are one and the
same. Corroborative evidence for this is given by calculations which
illustrate that the trace anomalies and Casimir energies obtained from the
two different descriptions are in agreement for known cases \cite%
{hensken,korea,adscftcalcs}. \ 

Generalizations of the counterterm action to asymptotically flat spacetimes
have also been proposed \cite{misner,Lau}. They are quite robust, and allow
for a full calculation of quasilocal conserved quantities in the Kerr
solution \cite{dehghani}\ that go well beyond the slow-rotating limit that
approximate embedding techniques require \cite{martinez}. \ Although they
can be inferred for general $d$ by considering spacetimes of special
symmetry, they cannot be algorithmically generated, and are in general
dependent upon the boundary topology \cite{KLS}.

Turning next to the asymptotically de Sitter case, we must add to the action
(\ref{action}) some counterterms to cancel its divergences

\begin{equation}
I_{ct}=\frac{1}{16\pi G}\int_{{\cal \partial M}^{+}}d^{d}x\sqrt{h}{\cal L}%
_{ct}+\frac{1}{16\pi G}\int_{{\cal \partial M}^{-}}d^{d}x\sqrt{h}{\cal L}%
_{ct}  \label{counter}
\end{equation}
so that 
\begin{equation}
I=I_{B}+I_{\partial B}+I_{ct}  \label{totaction}
\end{equation}
is now the total action.

For the special cases $d=2,3,4,$ the counterterm Lagrangian 
\begin{equation}
{\cal L}_{ct}=\gamma \left( -\frac{d-1}{\ell }+\frac{\ell \Theta \left(
d-3\right) }{2(d-2)}\hat{R}\right)  \label{count345}
\end{equation}
was proposed \cite{bala}, where $\hat{R}$ is the intrinsic curvature of the
boundary surfaces and the step function $\Theta (x)$ is equal to zero unless 
$x\geq 0$ which in this case it equals unity. The parameter $\gamma $ must
equal $-2\alpha $ to cancel divergences. The action (\ref{count345})\ \ was
shown to cancel divergences in de Sitter spacetime 
\begin{equation}
ds^{2}=-d\tau ^{2}+\ell ^{2}\exp \left( \tau ^{2}/\ell ^{2}\right) d\vec{x}%
\cdot d\vec{x}  \label{dsinf}
\end{equation}
and the Nariai spacetime 
\begin{equation}
ds^{2}=-\left( \frac{d\tau ^{2}}{\ell ^{2}}-1\right) ^{-1}d\tau ^{2}+\left( 
\frac{d\tau ^{2}}{\ell ^{2}}-1\right) dt^{2}+\ell ^{2}\left( 1-\frac{2}{d}%
\right) d\Omega _{d-1}^{2}  \label{Narai}
\end{equation}
where the metric $d\vec{x}\cdot d\vec{x}$ is a flat $d$-dimensional metric
that covers an inflationary patch of de Sitter spacetime and $d\Omega
_{d-1}^{2}$ is the metric of a unit $(d-1)$-sphere. Here 
\begin{equation}
\Lambda =\frac{d\left( d-1\right) }{2\ell ^{2}}  \label{cosmo}
\end{equation}
is the positive cosmological constant.

These results are suggestive that an algorithm similar to that in the AdS
case is applicable here, and indeed this is the case. \ Following the
procedure in ref. \cite{KLS}, we write the Einstein equations 
\begin{equation}
R_{\mu \nu }-\frac{1}{2}g_{\mu \nu }R=-\Lambda g_{\mu \nu }  \label{Ein}
\end{equation}
in the Gauss-Codacci form 
\begin{eqnarray}
\widehat{R}_{ab}-\frac{1}{2}\widehat{R}h_{ab}+u^{\mu }\nabla _{\mu }\Pi
_{ab}-\frac{1}{2}h_{ab}\left( \frac{\Pi ^{2}}{d-1}-\Pi _{cd}\Pi ^{cd}\right)
+\frac{\Pi }{d-1}\Pi _{ab} &=&\frac{d\left( d-1\right) }{2\ell ^{2}}h_{ab}
\label{gc1} \\
\nabla ^{b}\Pi _{ab} &=&0  \label{gc2} \\
\frac{1}{2}\left( \frac{\Pi ^{2}}{d-1}-\Pi _{cd}\Pi ^{cd}-\hat{R}\right) &=&%
\frac{d\left( d-1\right) }{2\ell ^{2}}  \label{gc3}
\end{eqnarray}
where $u_{\pm }^{\mu }$ is the timelike unit normal to ${\cal \partial M}%
^{\pm }$, whose metric is $h_{ab}^{\pm }$; eqs.(\ref{gc1}--\ref{gc3}) are
valid for each of these submanifolds. From the work of Mottola and Mazur %
\cite{Mozzola}, we know that the divergences of asymptotically de Sitter
spacetimes are independent of the boundary normal, and so depend only on
intrinsic boundary data. \ \ By writing the divergent part $\tilde{\Pi}_{ab}$
as a power series in $\ell $ 
\begin{equation}
\tilde{\Pi}_{ab}=\sum_{n=0}^{[d/2]}\ell ^{2n-1}\tilde{\Pi}_{ab}^{\left(
n\right) }  \label{lseries}
\end{equation}
it is easy to show that the trace $\tilde{\Pi}_{a}^{a\left( n\right) }$
appears linearly in eq. (\ref{gc3}), and so can be determined inductively in
terms all $\tilde{\Pi}_{ab}^{\left( k\right) }$, $k<n$, if these are known.
\ However these can be determined from the counterterm Lagrangian provided 
\begin{equation}
\tilde{\Pi}_{ab}=\frac{2}{\sqrt{h}}\frac{\delta }{\delta h_{ab}}\int d^{d}x%
\sqrt{h}{\cal L}_{ct}  \label{pivariation}
\end{equation}
so that under a Weyl rescaling $\delta _{W}h_{ab}=\sigma h_{ab}$\ we obtain
after some algebra 
\begin{equation}
\left( d-2n\right) {\cal L}_{ct}^{\left( n\right) }=\tilde{\Pi}_{a}^{a\left(
n\right) }  \label{weyl}
\end{equation}
up to an irrelevant total divergence, where ${\cal L}_{ct}=%
\sum_{n=0}^{[d/2]}\ell ^{2n-1}{\cal L}_{ct}^{(n)}$. \ \ 

The procedure for finding the counterterm Lagrangian for any given $d$ is
identical to the AdS case. Setting 
\begin{equation}
\tilde{\Pi}_{ab}^{\left( 0\right) }=(1-d)h_{ab}  \label{pitilde0}
\end{equation}
we obtain 
\begin{equation}
{\cal L}_{ct}^{\left( 0\right) }=\left( 1-d\right)  \label{Ltilde0}
\end{equation}
from (\ref{weyl}). Using this we insert the series (\ref{lseries}) into eq. (%
\ref{gc3}) and inductively obtain

\begin{eqnarray}
{\cal L}_{ct} &=&\left( -\frac{d-1}{\ell }+\frac{\ell \Theta \left(
d-3\right) }{2(d-2)}\hat{R}\right) -\frac{\ell ^{3}\Theta \left( d-4\right) 
}{2(d-2)^{2}(d-4)}\left( \hat{R}^{ab}\hat{R}_{ab}-\frac{d}{4(d-1)}\hat{R}%
^{2}\right)  \label{countergen} \\
&&-\frac{\ell ^{5}\Theta \left( d-5\right) }{(d-2)^{3}(d-4)(d-6)}\left( 
\frac{3d+2}{4(d-1)}\hat{R}\hat{R}^{ab}\hat{R}_{ab}-\frac{d(d+2)}{16(d-1)^{2}}%
\hat{R}^{3}-2\hat{R}^{ab}\hat{R}^{cd}\hat{R}_{acbd}\right.  \nonumber \\
&&\text{ \ \ \ \ \ \ \ \ \ \ \ \ \ \ \ \ \ \ \ \ \ \ \ \ \ \ \ \ \ \ \ }%
\left. -\frac{d}{4(d-1)}\nabla _{a}\hat{R}\nabla ^{a}\hat{R}+\nabla ^{c}\hat{%
R}^{ab}\nabla _{c}\hat{R}_{ab}\right)  \nonumber
\end{eqnarray}
for $d\leq 8$. \ The associated boundary stress-energy tensor can be
obtained by the variation of the action with respect to the variation of the
boundary metric, and is given by,

\begin{equation}
\begin{array}{c}
-8\pi GT_{ab}=\left( K_{ab}-Kh_{ab}\right) +\{\left( \frac{d-1}{\ell }h_{ab}+%
\frac{\ell \Theta \left( d-3\right) }{(d-2)}G_{ab}\right) \\ 
-\frac{\ell ^{3}\Theta \left( d-4\right) }{(d-2)^{2}(d-4)}\left[ -\frac{1}{2}%
h_{ab}\left( \hat{R}^{cd}\hat{R}_{cd}-\frac{d}{4(d-1)}\hat{R}^{2}\right) -%
\frac{d}{2(d-1)}\hat{R}\hat{R}_{ab}\right. \\ 
\left. -\frac{1}{2(d-1)}h_{ab}\nabla ^{2}\hat{R}+2\hat{R}^{cd}\hat{R}_{cadb}-%
\frac{d-2}{2(d-1)}\nabla _{a}\nabla _{b}\hat{R}+\nabla ^{2}\hat{R}_{ab}%
\right] \\ 
-\frac{2\ell ^{5}\Theta \left( d-5\right) }{(d-2)^{3}(d-4)(d-6)}\left\{ 
\frac{3d+2}{4(d-1)}\left( G_{ab}\hat{R}^{cd}\hat{R}_{cd}-\nabla _{a}\nabla
_{b}\left( \hat{R}^{ef}\hat{R}_{ef}\right) +h_{ab}\nabla ^{2}\left( \hat{R}%
^{ef}\hat{R}_{ef}\right) \right) +2\hat{R}\hat{R}_{a}^{\;c}\hat{R}%
_{bc}\right. \\ 
+h_{ab}\nabla _{c}\nabla _{d}\left( \hat{R}\hat{R}^{cd}\right) +\nabla
^{2}\left( \hat{R}\hat{R}_{ab}\right) -\nabla ^{c}\nabla _{b}\left( \hat{R}%
\hat{R}_{ac}\right) -\nabla ^{c}\nabla _{a}\left( \hat{R}\hat{R}_{bc}\right)
\\ 
-\frac{d(d+2)}{16(d-1)^{2}}\left[ -\frac{1}{2}h_{ab}\hat{R}^{3}+3\hat{R}^{2}%
\hat{R}_{ab}-3\nabla _{a}\nabla _{b}\hat{R}^{2}+3h_{ab}\nabla ^{2}\hat{R}^{2}%
\right] \\ 
-2\left[ -\frac{1}{2}h_{ab}\hat{R}^{ef}\hat{R}^{cd}\hat{R}_{ecfd}+\frac{3}{2}%
\left( \hat{R}_{a}^{\;e}\hat{R}^{cd}\hat{R}_{ecbd}+\hat{R}_{b}^{\;e}\hat{R}%
^{cd}\hat{R}_{ecad}\right) -\nabla _{c}\nabla _{d}\left( \hat{R}_{ab}\hat{R}%
^{cd}\right) +\nabla _{c}\nabla _{d}\left( \hat{R}_{a}^{\;c}\hat{R}%
_{b}^{\;d}\right) \right. \\ 
\left. +h_{ab}\nabla _{e}\nabla ^{f}\left( \hat{R}^{cd}\hat{R}%
_{\;cfd}^{e}\right) +\nabla ^{2}\left( \hat{R}^{cd}\hat{R}_{acbd}\right)
-\nabla _{e}\nabla _{a}\left( \hat{R}^{cd}\hat{R}_{\;cbd}^{e}\right) -\nabla
_{e}\nabla _{b}\left( \hat{R}^{cd}\hat{R}_{\;cad}^{e}\right) \right] \\ 
-\frac{d}{4(d-1)}\left[ \nabla _{a}\hat{R}\nabla _{b}\hat{R}-\frac{1}{2}%
h_{ab}\left( \nabla _{c}\hat{R}\nabla ^{c}\hat{R}\right) -2\hat{R}%
_{ab}\nabla ^{2}\hat{R}-2h_{ab}\nabla ^{4}\hat{R}+2\nabla _{a}\nabla
_{b}\nabla ^{2}\hat{R}\right] \\ 
+2\nabla _{c}\hat{R}_{ad}\nabla ^{c}\hat{R}_{b}^{\;d}+\nabla _{a}\hat{R}%
^{cd}\nabla _{b}\hat{R}_{cd}-\frac{1}{2}h_{ab}\nabla ^{e}\hat{R}^{cd}\nabla
_{e}\hat{R}_{cd} \\ 
-h_{ab}\nabla _{c}\nabla _{d}\nabla ^{2}\hat{R}^{cd}-\nabla ^{4}\hat{R}%
_{ab}+\nabla _{c}\nabla _{a}\nabla ^{2}\hat{R}_{b}^{\;c}+\nabla _{c}\nabla
_{b}\nabla ^{2}\hat{R}_{a}^{\;c} \\ 
-\nabla _{c}\left( \hat{R}_{bd}\nabla _{a}\hat{R}^{cd}\right) -\nabla
_{c}\left( \hat{R}_{ad}\nabla _{b}\hat{R}^{cd}\right) \\ 
\left. -\nabla _{c}\left( \hat{R}_{ad}\nabla ^{c}\hat{R}_{b}^{\;d}+\hat{R}%
_{bd}\nabla ^{c}\hat{R}_{a}^{\;d}\right) +\nabla ^{c}\left( \hat{R}%
_{cd}\nabla _{a}\hat{R}_{b}^{\;d}\right) +\nabla ^{c}\left( \hat{R}%
_{cd}\nabla _{b}\hat{R}_{a}^{\;d}\right) \right\} \}%
\end{array}
\label{stress}
\end{equation}

If the boundary geometry has an isometry generated by a Killing vector $\xi
^{\mu }$, then it is straightforward to show that $T_{ab}\xi ^{b}$ is
divergenceless. \ We write the boundary metric in the form 
\begin{equation}
h_{ab}d\hat{x}^{a}d\hat{x}^{b}=d\hat{s}^{2}=N_{t}^{2}dt^{2}+\sigma
_{ab}\left( d\varphi ^{a}+N^{a}dt\right) \left( d\varphi ^{b}+N^{b}dt\right)
\label{hmetric}
\end{equation}
where $\nabla _{\mu }t$ is a spacelike vector field that is the analytic
continuation of a timelike vector field and the $\varphi ^{a}$ are
coordinates describing closed surfaces $\Sigma $. From this it is
straightforward to show that the quantity 
\begin{equation}
{\frak Q}=\oint_{\Sigma }d^{d-1}\varphi \sqrt{\sigma }n^{a}T_{ab}\xi ^{b}
\label{Qcons}
\end{equation}
is conserved between surfaces of constant $t$, whose unit normal is given by 
$n^{a}$.\ Physically this would mean that a collection of observers on the
hypersurface whose metric is $h_{ab}$ would all observe the same value of $%
{\frak Q}$ provided this surface had an isometry generated by $\xi ^{b}$. \
If $\partial /\partial t$ is itself a Killing vector, then we can define 
\begin{equation}
{\frak M}=\oint_{\Sigma }d^{d-1}\varphi \sqrt{\sigma }N_{t}n^{a}n^{b}T_{ab}
\label{Mcons}
\end{equation}
as the conserved mass associated with the surface $\Sigma $\ at any given
point $t$ on the boundary. \ This quantity changes with the cosmological
time $\tau $. However a collection of observers that defined a surface $%
\Sigma $ would find that the value of ${\frak M}$ that they would measure
would not change as they collectively relocated to a different value of $t$
on the spacelike surface ${\cal \partial M}$. Since all asymptotically de
Sitter spacetimes must have an asymptotic isometry generated by $\partial
/\partial t$, there is at least the notion of a conserved total mass ${\frak %
M}$ for the spacetime as computed at future/past infinity. \ Similarly the
quantity 
\begin{equation}
{\frak J}^{a}=\oint_{\Sigma }d^{d-1}\varphi \sqrt{\sigma }\sigma
^{ab}n^{c}T_{bc}  \label{Jcons}
\end{equation}
can be regarded as a conserved angular momentum associated with the surface $%
\Sigma $\ if the surface has an isometry generated by $\partial /\partial
\phi ^{a}$.

{\bf \bigskip }

\section{Actions in de Sitter Spacetime}

\bigskip

We consider in this section an evaluation of the action using the
prescription (\ref{countergen}). From equation (\ref{Ein}), one gets 
\begin{equation}
R=-\Lambda \frac{d+1}{\left( 1-\left( d+1\right) /2\right) }=-\frac{d\left(
d-1\right) \left( d+1\right) }{\left( 1-d\right) \ell ^{2}}=\frac{d\left(
d+1\right) }{\ell ^{2}}  \label{ricciscalar}
\end{equation}
so that 
\begin{equation}
I_{B}=\frac{\alpha }{16\pi G}\int_{{\cal M}}d^{d+1}x\sqrt{-g}\left( \frac{%
d\left( d+1\right) }{\ell ^{2}}-\frac{d\left( d-1\right) }{\ell ^{2}}\right)
=\frac{d\alpha }{8\pi G\ell ^{2}}\int_{{\cal M}_{d}}d^{d}\hat{x}\int d\tau 
\sqrt{f}\sqrt{h}  \label{bulk2}
\end{equation}
where $V_{d}=\int_{{\cal M}_{d}}d^{d}\widehat{x}$ is the volume of the $d$
dimensional spatial section, and $\tau $ is the orthogonal coordinate
direction. \ 

The metric is of the form 
\begin{equation}
ds^{2}=-f\left( \tau \right) d\tau ^{2}+d\hat{s}^{2}  \label{genform}
\end{equation}
where $d\hat{s}^{2}$ is given by (\ref{hmetric}). Hence the timelike vector
normal to the hypersurface is 
\begin{equation}
u^{\mu }=\left( f^{-1/2}\left( \tau \right) ,0,0,\ldots ,0\right)
\label{normal}
\end{equation}
which yields 
\begin{equation}
\begin{array}{c}
K=h^{\mu \nu }\nabla _{\mu }u_{\nu }=-h^{\mu \nu }\Gamma _{\mu \nu
}^{\lambda }u_{\lambda }=-\frac{1}{2\sqrt{f}}h^{\mu \nu }\left( \partial
_{\nu }h_{\mu \tau }+\partial _{\mu }h_{\nu \tau }-\partial _{\tau }h_{\mu
\nu }\right) \\ 
=\frac{1}{2\sqrt{f}}h^{\mu \nu }\partial _{\tau }h_{\mu \nu }%
\end{array}
\label{ext}
\end{equation}
in turn giving 
\begin{equation}
I_{\partial B}=\frac{\beta }{32\pi G\sqrt{f}}\int_{{\cal M}_{d}^{-}}^{{\cal M%
}_{d}^{+}}d^{d}\widehat{x}\sqrt{h}\left( h^{\mu \nu }\partial _{\tau }h_{\mu
\nu }\right)  \label{bound}
\end{equation}
So we finally get 
\begin{equation}
\begin{array}{l}
I=I_{B}+I_{\partial B}+I_{ct} \\ 
=\frac{1}{16\pi G\ell ^{2}}\int_{{\cal M}_{d}^{-}}^{{\cal M}_{d}^{+}}d^{d}%
\hat{x}\left[ \left( 2d\alpha \int d\tau \sqrt{h}\sqrt{f}\right) +\sqrt{h}%
\left( \frac{\beta }{2\sqrt{f}}\left( h^{\mu \nu }\partial _{\tau }h_{\mu
\nu }\right) +{\cal L}_{ct}\right) \right]%
\end{array}
\label{tot}
\end{equation}
for the generic form of the action. We turn next to its evaluation in de
Sitter spacetime.

\subsection{Inflationary Coordinates}

The dS spacetime admits a coordinate system where equal time surfaces are
flat. In this case 
\begin{equation}
ds^{2}=-d\tau ^{2}+e^{2\tau /\ell }d\vec{x}^{2}  \label{bigbang}
\end{equation}
is the solution to the Einstein equations in de Sitter coordinates. $\tau $
changes from $-\infty $ to $+\infty ,$ and this patch (called the big bang
patch) covers half of the dS spacetime from a big bang at a past horizon to
the Euclidean surface at future infinity. The other half (big crunch patch)
of the dS spacetime from past infinity to a future horizon could be obtained
by replacing $\tau $ by $-\tau $ in (\ref{bigbang}). So, comparing with (\ref%
{genform}),\ $h_{\mu \nu }^{+}$ is a flat metric, and the counterterm
Lagrangian reduces to its first term for any $d$. Hence we have 
\begin{equation}
\begin{array}{c}
I=\frac{1}{16\pi G}\int_{{\cal M}_{d}^{+}}d^{d}\hat{x}\left[ 2d\alpha
\int_{-\infty }^{+\infty }\frac{d\tau }{\ell ^{2}}e^{d\tau /\ell }+\frac{%
\beta }{2}\left( 2d/\ell \right) \left. e^{d\tau /\ell }\right| _{-\infty
}^{+\infty }+\gamma \left. e^{d\tau /\ell }\right| _{\tau =+\infty }\left( 
\frac{1-d}{\ell }\right) \right] \\ 
=\frac{V_{d}}{16\pi G\ell }\left[ 2\alpha +\frac{\beta }{2}\left( 2d\right)
+\gamma \left( 1-d\right) \right] \left( \left. e^{d\tau /\ell }\right|
_{-\infty }^{\infty }\right)%
\end{array}
\label{tot2}
\end{equation}
where $V_{d}=\int_{{\cal M}_{d}^{+}}d^{d}\hat{x}$. This will diverge at $%
\tau =+\infty $ unless 
\begin{equation}
\beta =\gamma =-2\alpha  \label{para}
\end{equation}
in which case it vanishes. The quantity $V_{d}$ will also diverge unless the
boundary rendered compact, e.g. by toroidal identifications. \ 

So, for every dS spacetime in big bang coordinates the counterterm
Lagrangian (\ref{countergen}) removes all the divergences of the action (\ref%
{totaction}) in any dimension. A similar calculation shows that in the big
crunch coordinates with choosing (\ref{para}) the divergences of the action
are removed.

\subsection{Covering Coordinates}

Now we use global coordinates of the dS spacetime for which equal time
hypersurfaces are $d$-spheres $S^{d}$%
\begin{equation}
ds^{2}=-d\tau ^{2}+\ell ^{2}\cosh ^{2}\left( \tau /\ell \right) d\hat{\Omega}%
_{d}^{2}  \label{globall}
\end{equation}
\ These hypersurfaces have an infinitely large radius at $\tau =-\infty ,$
which decrease to a minimum value $\ell $ as $\tau \rightarrow 0$,
increasing again to infinity for $\tau =+\infty .$ Here $h_{\mu \nu }$ is
the metric of the $d$-sphere, so we have 
\begin{equation}
\begin{array}{c}
\hat{R}_{acbd}=\left( h_{ab}h_{cd}-h_{ad}h_{bc}\right) \text{ \ \ \ \ \ }%
\hat{R}_{ab}=\left( d-1\right) h_{ab}\text{ \ \ \ \ }\hat{R}=d\left(
d-1\right) \\ 
\hat{R}_{ab}\hat{R}^{ab}=\left( d-1\right) ^{2}d\text{ \ \ \ \ \ \ \ \ \ \ \
\ }\hat{R}^{ab}\hat{R}^{cd}\hat{R}_{acbd}=d\left( d-1\right) ^{3}%
\end{array}
\label{sphere}
\end{equation}
This in turn yields the following counterterm Lagrangian (\ref{countergen}) 
\begin{equation}
{\cal L}_{ct}=\gamma \left[ \left( -\frac{d-1}{\ell }+\frac{\Theta \left(
d-3\right) d\left( d-1\right) }{2\ell (d-2)\cosh ^{2}\left( \tau /\ell
\right) }\right) +\frac{\Theta \left( d-4\right) d(d-1)}{8\ell (d-4)\cosh
^{4}\left( \tau /\ell \right) }+\frac{\Theta \left( d-5\right) \left(
d-1\right) d}{16\ell (d-6)\cosh ^{6}\left( \tau /\ell \right) }\right]
\label{Lctsphere}
\end{equation}
and so the total action is 
\begin{equation}
\begin{array}{c}
I=\frac{\ell ^{d-1}V_{d}}{16\pi G}\left[ 2d\alpha \int_{-T}^{T}d\left( \tau
/\ell \right) \cosh ^{d}\left( \tau /\ell \right) +2\frac{\beta }{2}\left(
2d\right) \cosh ^{d-1}\left( T/\ell \right) \sinh \left( T/\ell \right)
\right. \\ 
+2\gamma \{\cosh ^{d}\left( T/\ell \right) \left( 1-d+\frac{\Theta \left(
d-3\right) d\left( d-1\right) }{2(d-2)\cosh ^{2}\left( T/\ell \right) }%
\right) +\frac{\Theta \left( d-4\right) d(d-1)\cosh ^{d-4}\left( T/\ell
\right) }{8(d-4)} \\ 
\left. +\frac{\Theta \left( d-5\right) \left( d-1\right) d\cosh ^{d-6}\left(
T/\ell \right) }{16(d-6)}\}\right]%
\end{array}
\label{action3}
\end{equation}
up to $d=8$. Using (\ref{para}), and setting \ $\alpha =1,$ we obtain

\begin{eqnarray}
I^{d=2} &=&\frac{\left( 4T/\ell +2\right) }{16\pi G}\ell V_{2}  \nonumber \\
I^{d=3} &=&0  \nonumber \\
I^{d=4} &=&\frac{\left( 6T/\ell -3/2\right) }{16\pi G}\ell ^{3}V_{4} 
\nonumber \\
I^{d=5} &=&0  \label{coveractions} \\
I^{d=6} &=&\frac{\left( 15T/2\ell -25/8\right) }{16\pi G}\ell ^{5}V_{6} 
\nonumber \\
I^{d=7} &=&0  \nonumber \\
I^{d=8} &=&\frac{\left( 35T/4\ell -413/96\right) }{16\pi G}\ell ^{7}V_{8} 
\nonumber
\end{eqnarray}
for the action (\ref{action3}) in the different dimensionalities.

We see that the action is finite up to a term that diverges linearly with $T$
\ for even $d$ as $T\rightarrow \infty $. This divergence cannot be removed
by a judicious choice of counterterms that are polynomials in boundary
curvature invariants because such invariants are all independent of $\ \tau $%
, as illustrated in eq. (\ref{sphere}). \ Clearly there are limitations to
the counterterm prescription. We shall comment on the implications of this
in terms of a possible dS/CFT correspondence in the concluding section.

\section{Schwarzchild-dS Spacetimes}

In this section, we consider the $d+1$ dimensional SdS spacetime. The metric
is 
\begin{equation}
ds^{2}=-N(r)dt^{2}+\frac{dr^{2}}{N(r)}+r^{2}d\hat{\Omega}_{d-1}^{2}
\label{Sdsmetric}
\end{equation}
where 
\begin{equation}
N(r)=1-\frac{2m}{r^{d-2}}-\frac{r^{2}}{\ell ^{2}}  \label{Sdslapse}
\end{equation}
and $d\hat{\Omega}_{d-1}^{2}$ denotes the metric on the unit sphere $%
S^{d-1}. $ For mass parameters $m$ with $0<m<m_{N}$, where 
\begin{equation}
m_{N}=\frac{\ell ^{d-2}}{d}(\frac{d-2}{d})^{\frac{d-2}{2}}  \label{naraimass}
\end{equation}
we have a black hole in dS spacetime with event horizon at $r=r_{H}$ and
cosmological horizon at $r=r_{C}>r_{H}.$ The event and cosmological horizons
locate in $N(r_{H})=N(r_{C})=0.$ When $m=m_{N},$ the event horizon coincides
with the cosmological horizon and one gets the Nariai solution. For $%
m>m_{N}, $ the metric (\ref{Sdsmetric}) describes a naked singularity in an
asymptotically dS spacetime. So demanding the absence of naked singularities
yields an upper limit to the mass of the SdS black hole. We want to work
outside of the cosmological horizon, where $N(r)<0$, \ so we set $r=\tau $
and rewrite the metric as 
\begin{equation}
ds^{2}=-f(\tau )d\tau ^{2}+\frac{dt^{2}}{f(\tau )}+\tau ^{2}d\tilde{\Omega}%
_{d-1}^{2}  \label{Sdsmet2}
\end{equation}
where 
\begin{equation}
f(\tau )=\left( \frac{\tau ^{2}}{\ell ^{2}}+\frac{2m}{\tau ^{d-2}}-1\right)
^{-1}  \label{flapse}
\end{equation}
The bulk action is now 
\begin{eqnarray}
I_{B} &=&\frac{d}{8\pi G\ell ^{2}}\int d^{d}x\int_{\tau _{+}}^{\tau }d\tau 
\sqrt{f}\sqrt{h}=\frac{d}{8\pi G\ell ^{2}}\int dtd^{d-1}\hat{x}\sqrt{\sigma }%
\int_{\tau _{+}}^{\tau }d\tau \tau ^{d-1}  \nonumber \\
&=&\frac{V_{d-1}^{t}}{8\pi G\ell ^{2}}\left( \tau ^{d}-\tau _{+}^{d}\right)
\label{sdsbulk}
\end{eqnarray}
where $V_{d-1}^{t}=\int dtd^{d-1}\hat{x}\sqrt{\sigma }$, where \ $\sigma
^{ab}$ is the metric on the unit $(d-1)$-sphere. Here $\tau _{+}$ is the
location of cosmological horizon which defined so that $\tau _{+}$ is the
largest root of $\ \left[ f(\tau _{+})\right] ^{-1}=0$; the integration is
from the cosmological horizon out to some fixed $\tau $ that will be sent to
infinity. We shall work in this ``upper patch'' outside of the cosmological
horizon in SdS spacetime; results for the lower patch are obtained in a
similar manner by setting $r=-\tau $ and considering $-\infty <\tau <-\tau
_{+}$ . \ 

The trace of the extrinsic curvature is 
\begin{eqnarray}
K &=&\frac{1}{2\sqrt{f}}h^{\mu \nu }\partial _{\tau }h_{\mu \nu }=\frac{%
\sqrt{f}}{2}\partial _{\tau }\left( \frac{1}{f}\right) +\frac{\left(
d-1\right) \tau ^{-2}}{2\sqrt{f}}\partial _{\tau }\tau ^{2}  \label{sdstrK}
\\
&=&\frac{1}{2\sqrt{f}}\left( -\frac{f^{\prime }}{f}+\frac{2\left( d-1\right) 
}{\tau }\right)  \nonumber
\end{eqnarray}
and so the boundary action becomes 
\begin{eqnarray}
I_{\partial B} &=&-\frac{1}{16\pi Gf}\int dtd^{d-1}\hat{x}\sqrt{\sigma }\tau
^{d-1}\left( -\frac{f^{\prime }}{f}+\frac{2\left( d-1\right) }{\tau }\right)
\nonumber \\
&=&-\frac{V_{d-1}^{t}}{16\pi Gf}\tau ^{d-1}\left( -\frac{f^{\prime }}{f}+%
\frac{2\left( d-1\right) }{\tau }\right)  \label{sdsbound}
\end{eqnarray}
where (\ref{para}) with \ $\alpha =1,$ has been employed.

Here $h_{\mu \nu }$ is the product metric of the $\left( d-1\right) $-sphere 
$\sigma _{ab}$ with $dt$, so we have 
\begin{eqnarray}
\hat{R}_{acbd} &=&\left( \sigma _{ab}\sigma _{cd}-\sigma _{ad}\sigma
_{bc}\right) \text{ \ \ \ \ \ }\hat{R}_{ab}=\left( d-2\right) \sigma _{ab}%
\text{ \ \ \ \ }\hat{R}=\left( d-2\right) \left( d-1\right)  \nonumber \\
\hat{R}_{ab}\hat{R}^{ab} &=&\left( d-2\right) ^{2}\left( d-1\right) \text{ \
\ \ \ \ \ \ \ \ \ \ \ }\hat{R}^{ab}\hat{R}^{cd}\hat{R}_{acbd}=\left(
d-1\right) \left( d-2\right) ^{3}  \label{sdsboundcurvs} \\
\nabla _{c}\hat{R}_{ab} &=&\nabla _{c}\hat{R}=0  \nonumber
\end{eqnarray}
where all $t$-components in any quantity in (\ref{sdsboundcurvs}) vanish.
Consequently 
\begin{eqnarray}
{\cal L}_{ct} &=&\left( -\frac{d-1}{\ell }+\frac{\ell \Theta \left(
d-3\right) }{2\tau ^{2}}\left( d-1\right) \right) -\frac{\ell ^{3}\Theta
\left( d-4\right) }{2(d-4)\tau ^{4}}\left( d-1\right) \left( 1-\frac{d}{4}%
\right)  \nonumber \\
&&-\frac{\ell ^{5}\Theta \left( d-5\right) }{(d-4)(d-6)\tau ^{6}}\left(
d-1\right) \left[ \frac{3d+2}{4}-\frac{d(d+2)}{16}-2\right]  \label{sdsLct}
\\
&=&\left( \frac{1-d}{\ell }+\frac{\ell \Theta \left( d-3\right) }{2\tau ^{2}}%
\left( d-1\right) \right) +\frac{\ell ^{3}\Theta \left( d-4\right) }{8\tau
^{4}}\left( d-1\right) +\frac{\ell ^{5}\Theta \left( d-5\right) }{16\tau ^{6}%
}\left( d-1\right)  \nonumber
\end{eqnarray}

So using (\ref{para}) the action becomes 
\begin{equation}
\begin{array}{c}
I=\frac{V_{d-1}^{t}}{8\pi G\ell ^{2}}\left( \tau ^{d}-\tau _{+}^{d}\right) -%
\frac{V_{d-1}^{t}}{16\pi Gf}\tau ^{d-1}\left( -\frac{f^{\prime }}{f}+\frac{%
2\left( d-1\right) }{\tau }\right) \\ 
-\frac{\left( d-1\right) V_{d-1}^{t}\tau ^{d-1}}{8\pi G\sqrt{f}}\left[
\left( -\frac{1}{\ell }+\frac{\ell \Theta \left( d-3\right) }{2\tau ^{2}}%
\right) +\frac{\ell ^{3}\Theta \left( d-4\right) }{8\tau ^{4}}+\frac{\ell
^{5}\Theta \left( d-5\right) }{16\tau ^{6}}\right] \\ 
=\frac{V_{d-1}^{t}\tau ^{d-1}}{8\pi G}\left[ \frac{1}{\ell ^{2}}(\tau -\frac{%
\tau _{+}^{d}}{\tau ^{d-1}})-\frac{1}{2f}\left( -\frac{f^{\prime }}{f}+\frac{%
2\left( d-1\right) }{\tau }\right) -\frac{\left( d-1\right) }{\sqrt{f}}\left[
\left( -\frac{1}{\ell }+\frac{\ell \Theta \left( d-3\right) }{2\tau ^{2}}%
\right) +\frac{\ell ^{3}\Theta \left( d-4\right) }{8\tau ^{4}}+\frac{\ell
^{5}\Theta \left( d-5\right) }{16\tau ^{6}}\right] \right]%
\end{array}
\label{action4}
\end{equation}%
and we obtain the following form of the actions in different dimensions%
\begin{eqnarray}
I^{d=2} &=&-\frac{\left( m-1/2+\tau _{+}^{2}/\ell ^{2}\right) V_{1}^{t}}{%
8\pi G}  \nonumber \\
I^{d=3} &=&-\frac{\left( m+\tau _{+}^{3}/\ell ^{2}\right) V_{2}^{t}}{8\pi G}
\nonumber \\
I^{d=4} &=&-\frac{\left( m-3\ell ^{2}/8+\tau _{+}^{4}/\ell ^{2}\right)
V_{3}^{t}}{8\pi G}  \nonumber \\
I^{d=5} &=&-\frac{\left( m+\tau _{+}^{5}/\ell ^{2}\right) V_{4}^{t}}{8\pi G}
\label{sdsactions} \\
I^{d=6} &=&-\frac{\left( m-5\ell ^{4}/16+\tau _{+}^{6}/\ell ^{2}\right)
V_{5}^{t}}{8\pi G}  \nonumber \\
I^{d=7} &=&-\frac{\left( m+\tau _{+}^{7}/\ell ^{2}\right) V_{6}^{t}}{8\pi G}
\nonumber \\
I^{d=8} &=&-\frac{\left( m-35\ell ^{6}/128+\tau _{+}^{8}/\ell ^{2}\right)
V_{7}^{t}}{8\pi G}  \nonumber
\end{eqnarray}%
\ in the limit $\tau \rightarrow +\infty $. We note that all actions are
finite. Note that the $\tau _{+}$--independent terms for even $d$ are
consistently positive.

From (\ref{Mcons}) the mass is 
\begin{equation}
{\frak M}=\int d^{d-1}\hat{x}\sqrt{\sigma }\tau ^{d-1}N_{t}n^{a}n^{b}T_{ab}=%
\sqrt{f}\int d^{d-1}\hat{x}\sqrt{\sigma }\tau ^{d-1}T_{tt}  \label{sdsmass}
\end{equation}
where 
\begin{equation}
n^{a}=(0,\sqrt{f},\vec{0})  \label{tnorm}
\end{equation}
is the unit normal in the $t$-direction and $N_{t}=\frac{1}{\sqrt{f}}$. \
The extrinsic curvature $K_{ab}=h_{a}^{\mu }\nabla _{\mu }u_{b}$ is 
\begin{eqnarray}
K_{tt} &=&h_{t}^{\mu }\nabla _{\mu }u_{t}=h_{t}^{\mu }\left( \partial _{\mu
}u_{t}-\Gamma _{\mu t}^{\lambda }u_{\lambda }\right) =-\frac{1}{2\sqrt{f}}%
\left( 2\partial _{t}h_{t\tau }-\partial _{\tau }h_{tt}\right)  \nonumber \\
&=&\frac{1}{2\sqrt{f}}\partial _{\tau }h_{tt}=-\frac{f^{\prime }}{2f^{2}%
\sqrt{f}}  \label{Kext}
\end{eqnarray}
where the prime refers to the derivative with respect to $\tau $. \ 

\ Since there is constant curvature in the $(d-2)$-dimensional subspace and
all $t$-components vanish from the curvatures in (\ref{sdsboundcurvs}), we
have

\begin{equation}
\begin{array}{c}
T_{tt}=\frac{1}{4\pi G}[\left( K_{tt}-Kh_{tt}\right) +\left( \frac{d-1}{\ell 
}h_{tt}-\frac{\ell \Theta \left( d-3\right) }{2(d-2)}h_{tt}\hat{R}\right) -%
\frac{\ell ^{3}\Theta \left( d-4\right) }{(d-2)^{2}(d-4)}\left\{ -\frac{1}{2}%
h_{tt}\left( \hat{R}^{cd}\hat{R}_{cd}-\frac{d}{4(d-1)}\hat{R}^{2}\right)
\right\} \\ 
-\frac{2\ell ^{5}\Theta \left( d-5\right) }{(d-2)^{3}(d-4)(d-6)}\{\frac{3d+2%
}{4(d-1)}\left( -\frac{1}{2}h_{tt}\hat{R}\hat{R}^{cd}\hat{R}_{cd}\right) -%
\frac{d(d+2)}{16(d-1)^{2}}\left[ -\frac{1}{2}h_{tt}\hat{R}^{3}\right] -2-%
\frac{1}{2}h_{tt}\hat{R}^{ef}\hat{R}^{cd}\hat{R}_{ecfd}\text{\ } \\ 
\text{\ }+h_{tt}\nabla _{e}\nabla ^{f}\left( \hat{R}^{cd}\hat{R}%
_{\;cfd}^{e}\right) +\nabla ^{2}\left( \hat{R}^{cd}\hat{R}_{tctd}\right)
-2\nabla _{e}\nabla _{t}\left( \hat{R}^{cd}\hat{R}_{\;ctd}^{e}\right) \}]%
\end{array}
\label{stress2}
\end{equation}
or

\begin{equation}
\begin{array}{c}
T_{tt}=\frac{1}{4\pi G}[-\frac{1}{2f\sqrt{f}}\left( \frac{2\left( d-1\right) 
}{\tau }\right) +\left( \frac{d-1}{f\ell }-\frac{\ell \Theta \left(
d-3\right) }{2f\tau ^{2}}\left( d-1\right) \right) \\ 
-\frac{\ell ^{3}\Theta \left( d-4\right) }{8f\tau ^{4}}\left( d-1\right) -%
\frac{2\ell ^{5}\Theta \left( d-5\right) }{32f\tau ^{6}}(d-1)]%
\end{array}
\label{stress3}
\end{equation}%
From this component of stress tensor, using (\ref{sdsmass}), we obtain 
\begin{equation}
\begin{array}{l}
{\frak M}^{d=2}=-\frac{V_{1}}{16\pi G}(2m-1)\{1+\frac{1}{4\tau ^{2}}\ell
^{2}(2m-1)+{\cal O}(\frac{1}{\tau ^{4}})\} \\ 
{\frak M}^{d=3}=-\frac{V_{2}}{16\pi G}\left( 4m-\frac{\ell ^{2}}{2\tau }+%
{\cal O}(\frac{1}{\tau ^{3}})\right) \\ 
{\frak M}^{d=4}=-\frac{V_{3}}{16\pi G}\left( 6m-\frac{3}{4}\ell ^{2}+\frac{%
3\ell ^{2}}{64\tau ^{4}}(8m-\ell ^{2})^{2}+{\cal O}(\frac{1}{\tau ^{6}}%
)\right) \\ 
{\frak M}^{d=5}=-\frac{V_{4}}{16\pi G}\left( 8m-\frac{\ell ^{4}}{2\tau }+%
{\cal O}(\frac{1}{\tau ^{3}})\right) \\ 
{\frak M}^{d=6}=-\frac{V_{5}}{16\pi G}(10m-\frac{5}{8}\ell ^{4}-\frac{5\ell
^{6}}{64\tau ^{2}}+{\cal O}(\frac{1}{\tau ^{6}})) \\ 
{\frak M}^{d=7}=-\frac{V_{6}}{16\pi G}\left( 12m-\frac{15}{32}\frac{\ell ^{6}%
}{\tau }+{\cal O}(\frac{1}{\tau ^{3}})\right) \\ 
{\frak M}^{d=8}=-\frac{V_{7}}{16\pi G}\left( 14m-\frac{35}{64}\ell ^{6}-%
\frac{7}{64}\frac{\ell ^{8}}{\tau ^{2}}+{\cal O}(\frac{1}{\tau ^{4}})\right)%
\end{array}
\label{masses}
\end{equation}%
where we have retained the leading terms in $\tau $ in the large-$\tau $
limit. \ For odd values of $d$,\ ${\frak M}$ \ is an increasingly negative
function of $\tau $, approaching a constant negative value as $\tau
\rightarrow \infty $.\ For even values of $\ d$ the situation is reversed:\ $%
{\frak M}$ \ is an increasingly positive function of $\tau $, approaching a
constant positive value as $\tau \rightarrow \infty $. As the mass parameter 
$m$ increases, this constant positive value decreases, approaching its
minimum at the Nariai limit.\ Setting $m=0$ gives the mass of dS spacetime
in different dimensionalities. We note that dS spacetime with even
dimensions has zero mass and the others with even dimensions have positive
mass. Our results in the special case of $\ d=4$ agrees with the known
Casimir energy of the dual CFT living on the boundary of dS$_{5}$ \cite{Klem}
(up to a sign, because our signs are the same as \cite{bala}). Also, we note
that our mass formula in the special case of \ $d=2$ agrees with the result
of \cite{Myung}. Furthermore, we observe that if the dual CFT theory exists %
\cite{Witten, Stro} (like the CFT dual to AdS spacetime), then the dual of
the calculated mass ${\frak M}^{d}$ (\ref{masses}) is the energy of a
boundary Euclidean CFT$_{d}$. Since our mass ${\frak M}^{d}{\frak \ }$%
decreases with increasing black hole mass parameter $m$, so the entropy of
the dual boundary theory which is proportional to the energy of the boundary
CFT, decreases relative to its de Sitter maximum.

The volumes $V_{d-1}^{t}$ are in general divergent, since the $t$-coordinate
is of infinite range. However since $\partial /\partial t$ is a Killing
vector, it is tempting to periodically identify it. \ Indeed, if we
analytically continue $t\rightarrow it$, we obtain a metric of signature $%
(-2,d-1)$. \ The submanifold of signature $\left( -,-\right) $ described by
the $(t,\tau )$ coordinates will have a conical singularity at $\tau =\tau
_{+}$ unless the $t$-coordinate is periodically identified with period 
\begin{equation}
\beta _{H}=\left| \frac{4\pi }{(-N^{\prime }(r))}\right| _{r=\tau
_{+}}=\left| \frac{-f^{\prime }(\tau )}{4\pi f^{2}(\tau )}\right| _{\tau
=\tau _{+}}^{-1}  \label{betaH}
\end{equation}%
This is the analogue of the Hawking temperature outside of the cosmological
horizon. \ 

Proceeding further, we can provisionally define an `entropy' by analytically
continuing the gravitational Gibbs-Duhem relation \cite{TNUT}:

\begin{equation}
S=\beta _{H}{\frak M}_{\tau \rightarrow \infty }-I  \label{entropy}
\end{equation}%
where $\beta _{H}=\oint dt$ is the Euclideanized integral over $t.$This
gives 
\begin{equation}
S_{d}=\frac{\left( \tau _{+}^{d}-(d-2)m\ell ^{2}\right) \beta _{H}V_{d-1}}{%
8\pi G\ell ^{2}}  \label{sdsentropy}
\end{equation}%
up to $d=8$. \ It is straightforward to show that these entropies are always
positive, since $\tau _{+}^{d}>(d-2)m\ell ^{2}$ so long as $m<m_{N}$.

For example, for $d=2$, we have 
\begin{equation}
\begin{array}{c}
\frac{\tau _{+}^{2}}{\ell ^{2}}+2m-1=0\Rightarrow \tau _{+}=\ell \sqrt{1-2m}
\\ 
\Rightarrow S^{d=2}=\frac{\left( 1-2m\right) \beta _{H}V_{1}}{8\pi G}%
\end{array}
\label{two}
\end{equation}
and from (\ref{betaH}) we have $\beta _{H}=2\pi \ell ^{2}/\tau _{+}$, so 
\begin{equation}
S^{d=2}=\frac{\tau _{+}V_{1}}{4G}=\frac{\pi \ell \sqrt{1-2m}}{2G}
\label{d2entropy}
\end{equation}
in agreement with ref. \cite{bala} and \cite{Myung}, provided we set $%
1-2m\rightarrow M$, which is the metric for a conical deficit. Moreover,
from expressions (\ref{sdsentropy}), we see that in any dimension up to
nine, the entropy of SdS black hole is monotonically decreasing function of
the mass parameter. Hence the entropy of a massive SdS black hole is less
than the entropy of empty dS spacetime. So, the D-bound \cite{bousso} on the
entropy of asymptotically dS spacetimes with positive cosmological constant
is satisfied.

\section{Discussion}

We have shown in this paper that the counterterm generating algorithm in
asymptotically AdS spacetimes \cite{KLS}\ can be generalized to the
asymptotically de Sitter case. \ We have explicitly computed the counterterm
Lagrangian and associated boundary stress-energy up to $d=8$. \ The results
are a straightforward analytic continuation of the anti de Sitter case.
However their interpretation is somewhat less clear. \ While the conserved
charges (\ref{Qcons}) can be associated with the closed surface $\Sigma $,
this surface does not itself enclose anything, since the topology of the
hypersurface is ${\Bbb R}\times S^{d-1}$. \ Consequently one cannot really
consider the conserved charges as being contained within $\Sigma $. \
However this is not as dissimilar to the asymptotically flat and anti de
Sitter cases (where $\partial /\partial t$ is timelike)\ as it may first
appear. In those situations, the conserved quantities are defined on a given
spatial slice which evolves in time. \ Although it is natural to think of
these quantities as being contained within the hypersurface this need not be
so, since anything that the observers measure are contingent only upon their
choice of closed spatial surface on a given spacelike slice, and not on
anything that takes place within the surface \cite{nonortho}. Indeed, they
need not even be aware than an interior exists! \ 

Notwithstanding such interpretive subtleties, the results are tantalizingly
similar to those obtained in the AdS case, and provide further evidence for
a possible dS/CFT correspondence. \ The terms proportional to $\ell ^{d-2}$
in the action and in the conserved masses for even $d$ are presumably the
analogues of the Casimir energy of the dual Euclidean CFT, and have been
shown to be such in restrictive cases \cite{Klem}. The masses are
consistently negative in any dimension, and the additional contributions in
even $d$ are consistently positive. Their AdS counterparts could be obtained
by a Wick rotation $\ell \rightarrow i\ell $; indeed the dS procedure is in
some sense a ``Wick rotation'' of the AdS procedure \cite{mcinnes}. \ 

However note that the mass parameter $m$ can be negative; although the
spacetime has singularities, these are always hidden from observers outside
of the cosmological horizon. \ For observers located in the ``lower patch''
\ of the \ Penrose diagram these singularities are hidden behind a future
horizon, whereas they are in behind the past horizon for observers in the
``upper patch''. \ If negative values of $m$ are permitted then the
conserved mass ${\frak M}$ \ is always positive and greater than its value
in de Sitter spacetime; moreover observers outside the cosmological horizon
will never encounter the singularities. Whether or not this violates the
conjecture of ref. \cite{bala} will depend upon a clarification of the
notion of a cosmological singularity in this context. \ 

We also found that the counterterm Lagrangian cannot always cancel
divergences in the action. Specifically, for de Sitter spacetime there are
divergences in the action for even values of $d$ when the boundary geometry
is $S^{d}$. \ \ These divergences are the de Sitter analogues of those found
in the AdS case \cite{EJM} for compact boundary geometries of the form $%
S^{d} $ or $H^{d}$, where the latter is a compact hyperbolic space of
non-trivial topology. \ For reasons similar to the AdS case, this need not
be fatal to a putative dS/CFT correspondence conjecture -- the linear
divergence could be reflective of a UV divergence in the Euclidean\ CFT. \ 

Finally, we constructed a provisional definition of entropy by generalizing
the Gibbs-Duhem relation in asymptotically flat and AdS spacetimes. By
analytically continuing the $t$-coordinate to imaginary values {\it outside}
of the cosmological horizon, we find that the resultant metric will have
conical singularities unless the imaginary $t$-coordinate is appropriately
periodically identified. Although the choice of periodicity needs to be more
fully justified, \ it yields finite and well-defined values for the
provisional entropy (\ref{entropy}) which are in agreement with those
obtained from CFT methods in $d=2$ and which satisfy the D-bound on the
entropy of asymptotically dS spacetimes. This suggests that a well-defined
notion of gravitational entropy outside of a cosmological horizon can be
meaningfully constructed. Its full meaning within the context of
gravitational thermodynamics remains a subject for future investigation.

\bigskip

{\Large Acknowledgments}

This work was supported by the Natural Sciences and Engineering Research
Council of Canada. We would like to thank D. Marolf and R. Myers for
discussions.

\end{document}